\documentclass[a4paper,twoside,11pt,fullpage]{llncs}
\usepackage{amsfonts}
\usepackage{amsmath}
\usepackage{graphicx}
\usepackage{algorithm}
\usepackage[noend]{algorithmic}

\newcommand{\dhop}{\ensuremath{d^{hop}}} 
\newcommand{\da}{\ensuremath{d^{anchor}}} 
\newcommand{\did}{\ensuremath{\overrightarrow {d_{ideal}}}}
\newcommand{\ddir}{\ensuremath{\overrightarrow {d_{direction}}}}
\newcommand{\dlast}{\ensuremath{\overrightarrow {d_{last}}}}
\newcommand{\vGamma}{\ensuremath{\overrightarrow {\Gamma}}}
\newcommand{\north}{\ensuremath{\overrightarrow {v_{north}}}}
\newcommand{\east}{\ensuremath{\overrightarrow {v_{east}}}}
\newcommand{\peast}{\ensuremath{\overrightarrow {p_{east}}}}

\title{VRAC: Simulation Results \#1}

\author{Florian Huc and Aubin Jarry}

\institute{Computer Science Department \\University of Geneva\\1211 Geneva 4,Switzerland\\ firstname.lastname@unige.ch }

\begin{document}
\maketitle

\subsection*{Abstract}

In order to make full use of geographic routing techniques developed for large scale networks, nodes must be localized. However, localization and virtual localization techniques in sensor networks are dependent either on expensive and sometimes unavailable hardware (e.g. GPS) or on sophisticated localization calculus (e.g. triangulation) which are both error-prone and with a costly overhead.

Instead of localizing nodes in a traditional 2-dimensional space, we use directly the raw distance to a set of anchors to route messages in a multi-dimensional space. This should enable us to use any geographic routing protocol in a robust and efficient manner in a very large range of scenarios. We test this technique for two different geographic routing algorithms, namely GRIC and ROAM. The simulation results show that using the raw coordinates does not decrease their efficiency.

\vspace{3cm}

\keywordname { Sensor Networks, Routing, Anchors, Virtual Coordinates}

\newpage
\section{Introduction, State of the Art}

Sensor networks can be used in a wide range of applications. A prominent application is to make measurements on a wide area and to gather all this measures at one or several sinks. One can also consider scenarios in which any pair of sensors may want to communicate \cite{AI+02}.

Although, in wired networks each node is equipped with substantial computation and storage resources, and can maintain routing tables, this is far too costly in sensor networks which are made of small and cheap devices. Indeed, the computation phase requires energy (which is a limited resource) and the storage of data requires energy and space (also a limited resource). Instead of using routing tables, local routing techniques have been developed. A compelling technique consists in using nodes' coordinates. Many algorithms have been devised such as GPSR \cite{KK00} and  OAFR \cite{KWZ08} which use greedy routing and face routing on a planarized connectivity graph. One can also cite GRIC \cite{PS07} a greedy routing algorithm following the sides of an obstacle when one is met, and which introduces some inertia in the direction followed by the message. If one authorizes the use of a bit of memory at each node, then early obstacle detection algorithms have been proposed \cite{ML+08,HJM+09}.

To obtain coordinates, we may suppose that each sensor is equipped with a GPS.
 However, the hypothesis of having a GPS for each sensor arguably leads to too expensive devices. This assumption may be weakened by equipping a subset of the sensors with GPS; these sensors are usually called anchors. From them, approximate coordinates can be computed for all nodes of the network. In \cite{LR03}, three such algorithms are compared, namely Ad-hoc positioning, Robust positioning, and N-hop multilateration. One can also cite the algorithm At-Dist \cite{BKS08}, which is a distributed algorithm estimating the position of each node together with an estimate of its accuracy. Some authors, improved these results by using angles measurement \cite{BGJ05}. However, note that these techniques need flooding from anchors and many computations at each node. Hence they are energy consuming. Furthermore, the computed coordinates are approximations which turns to be often insufficient. Also note that angle measurement needs potentially costly extra devices. Other papers \cite{MLRT04,MobileAssist_INFOCOM2005,CC+05,RPSS03} propose to compute virtual coordinates from the distance between nearby nodes and have the advantage of not needing anchors. These approaches have two aspects: first to determine potential coordinates, and secondly to determine if they are unique or not. In \cite{MobileAssist_INFOCOM2005}, the authors use a mobile unit to assist in measuring the distance between nodes in order to improve accuracy. The algorithm proposed in \cite{CC+05} first chooses three nodes that will behave as anchors and from which virtual coordinates are determined. If these techniques do not need any GPS since there are no anchors, they suffer from their inaccuracy or high energy consumption in a preprocessing phase.

 In this paper, we want to avoid any preprocessing technique and propose to directly use raw information. We consider that there are some special nodes in the network, to which any other node knows its distance (cf Section \ref{sec:simul} to see how such datas are supposed to be known or computed). By similarity, we call them anchors even if they are not equipped with GPS. Our objective is to study routing techniques using directly the distance to the anchors as coordinates, without computing from them 2-dimensional coordinates. Eventually, we want to adapt this idea to any algorithm using geographic techniques, for a start, we experimented it with GRIC~\cite{PS07} and ROAM~\cite{HJM+09}.

\section{Concept}
\subsection{Coordinates}
Given a node at location $X$, we define the multi-dimensional coordinates $f(X)$ of this node as its distance to some anchors at location $A_1, A_2, \dots A_n$:

$$f:X\rightarrow \begin{pmatrix} d(X,A_1) \\ d(X,A_2) \\ \dots \\ d(X,A_n) \end{pmatrix}.$$

Observe that any continuous distance $d$ on the plane will produce a continuous surface in $\mathbb{R}^n$ and that any distance function $d$ may be used. In this paper, we consider both the Euclidean distance and the distance in terms of hops.

\paragraph{Euclidean distance}
In the plane with Euclidean distance, a node at location $X$ has a pair of absolute coordinates $(x,y)$. If the anchor at location $A_i$ has for absolute coordinates $(x_i, y_i)$ then $f$ is a function from $\mathbb{R}^2\rightarrow\mathbb{R}^n$ defined by
$$f:(x,y)\rightarrow \begin{pmatrix} \sqrt{(x-x_1)^2 + (y-y_1)^2} \\ \sqrt{(x-x_2)^2 + (y-y_2)^2} \\ \dots \\ \sqrt{(x-x_n)^2 + (y-y_n)^2} \end{pmatrix}.$$
Since the functions $f_i:(x,y)\rightarrow\sqrt{(x-x_i)^2 + (y-y_i)^2}$ are continuous and in $C^\infty$ except in $(x_i,y_i)$, the image $f(\mathbb{R}^2)$ in $\mathbb{R}^n$ is a {\em continuous surface} with singularities at the image of anchors.

\begin{figure}[tb]
 \begin{center}
  \includegraphics[width=10cm]{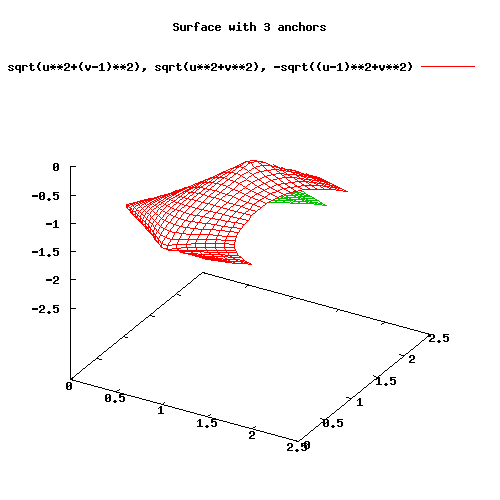}
  \caption{Representation of the distance to three anchors.}
  \label{fig:3a}
 \end{center}
\end{figure}

Figure \ref{fig:3a} represents the image of $f$, when there are three anchors at location $(0,0)$, $(0,1)$ and $(1,0)$.

\paragraph{Hop distance}
The distance in terms of hops between two nodes $u$ and $v$ is the minimum number $k$ of edges in a path from $u$ to $v$: $u,v_2,\dots,v_{k-1},v$. We note it $\dhop(u,v)=k$.

This defines correctly a distance function since one can easily verify that:
\begin{itemize}
\item for all pairs of nodes $u,v$, we have $d^hop(u,v)=0$ iff $u=v$
\item given three nodes $u,v,w$, $\dhop(u,v)\leq \dhop(u,w)+\dhop(w,v)$
\item the hop distance is symmetric and positive.
\end{itemize}

\subsection{Anchors distance}
From the function $f$, we define a new distance function which we call \emph{anchors distance}:
given two nodes $u,v$, $\da(u,v)=||f(u)-f(v)||_2=\sqrt{\sum_{i=1}^n(f_i(u)-f_i(v))^2}$

To check that $\da$ is a distance function, notice that:
\begin{itemize}
  \item $\da$ is positive.
  \item  $\da$ satisfies the triangular inequality. Indeed, for any $1 \leq i \leq n$ and any three nodes $u,v,w$, $\sqrt{f_i(u)^2-f_i(v)^2}\leq \sqrt{f_i(u)^2-f_i(w)^2}+\sqrt{f_i(w)^2-f_i(v)^2}$.
  \item $\da$ is symmetric. 
\end{itemize}

It only remains to check that when $\da(u,v)=0 \Rightarrow u=v$. Indeed, it is not always true, in particular if there are one or two anchors, two nodes in a plane may satisfy $\da(u,v)=0$ and $u \neq v$. This question is answered by Lemma~\ref{lem:distance} for the case of the Euclidean distance. In the case of hopdistance the problem was studied in \cite{CC+05}. The authors show that with anchors forming an equilateral triangle, nodes that share the same coordinates inside the triangle form ambiguity zones whose diameters are bounded by the communication radius of nodes at high density.

\begin{definition}
Given a $k$ dimensional space, we say that $k+1$ points are in a general setting if no $k-1$ subspace contains all points.
\end{definition}

\begin{lemma}\label{lem:distance} 
  Given a $k$ dimensional space with the Euclidean distance $d$, if there are $k+1$ anchors in a general setting, then $\da$ is a distance.
\end{lemma}

\begin{proof}
We need to check that given two nodes $u,v$ of a $k$ dimensional space, $\da(u,v)=0 \Rightarrow u=v$ if we have at least $k+1$ anchors in a general setting.

If $u$ and $v$ are two different points, then the subspace of points equidistant to them defined as  $\{x:d(u,x)=d(v,x)\}$ is a hyperplane, hence it cannot contain all anchors by hypothesis. So we have $u=v$.
\end{proof}

\begin{remark}
  If $\da$ is defined using $\dhop$, the previous lemma may not be verified. Indeed, two points which are close one to another may have exactly the same coordinates in terms of hop. As long as they are neighbors one of another, this is not a problem. Giving them a different identity makes it possible to recognize them when the message reaches its destination in terms of coordinates. This problem did not occur during the experiments.
\end{remark}

\subsection{Difficulties}

\begin{figure}[tb]
 \begin{center}
  \includegraphics[width=5cm]{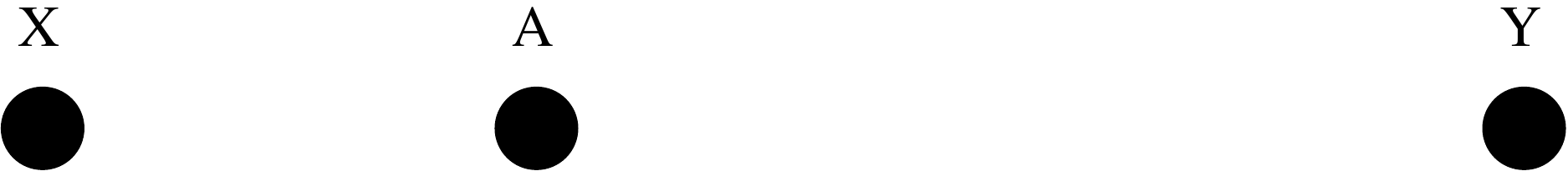}
  \caption{Anchor in the middle problem}
  \label{fig:AnchorMidle}
 \end{center}
\end{figure}

\paragraph{Anchor in the middle problem}
The surface singularities near anchors is a consequence of the more general {\em anchor-in-the-middle} problem. Suppose that there are three locations $A, X, Y$ in the Euclidean plane such that $\vec{XY} = 3\times\vec{XA}$ (see Figure~\ref{fig:AnchorMidle}). While trying to route from point $X$ to point $Y$, a coordinate related to $A$ will tell $X$ to send the message away from $A$, since $X$ is closer to $A$ then $Y$, which is exactly the wrong thing to do in this context.
This problem is independent from the choice of the distance function, but is not without solutions. An escape solution would be to have only anchors on the boundary of the network. This is a common strategy with virtual coordinates, used for instance in \cite{CC+05,RPSS03}. A second solution would be to have enough anchors so that the problem is mitigated. A third solution would be to wisely select anchors while routing, ignoring anchors whose distance to the sender is less than the distance to the destination and less than half the distance to the farthest anchor. Nevertheless, our experiments show that choosing only six anchors at random in the whole network is sufficient to solve the problem.

\paragraph{Curve}
The surface $f(\mathbb{R}^2)$ has a {\em curve}. If we use the Euclidean distance, the surface will be almost flat far away from the anchors, but the curve will be more pronounced near the anchors (see Figure~\ref{fig:3a}). The curve will probably lead to somewhat sub-optimal curved paths in the original $\mathbb{R}^2$, however it does not prevent from finding a path. As an illustration, ships do not follow constant latitude paths, which may be up to $\frac{\pi}{2}$ times longer than wanted near the north or south pole. This means that some geographic properties may not be preserved, however properties such as connectivity and subgraphs properties are preserved. 

\paragraph{Computation cost}
While saving on initialization overhead, multi-dimensional routing causes some additional computation costs when sending messages. First, we point out that these computation costs are not communication costs and should be lower in terms of energy consumption by some order of magnitude. Then, we believe that the added costs are most likely that of multi-dimensional scalar products, which are just some added additions and multiplications, while the expensive operations (square root extraction, divisions) stay the same as in traditional 2-dimensional routing.

\subsection{Scenarios}
We propose several scenarios for which our technique should be used. There are two main types. In the first one, the anchors are special nodes of the network which can advertise their distance in hop to the other nodes, for example via a flooding. We suppose that this nodes are chosen at random between all nodes of the network.

The second type of scenario is when the signal is emitted by some external entities, as a plane or a robot which deploy the network. Just after it would deliver a powerful signal at different positions which play the role of anchors. Another option is to have special devices, which can be seen as signal bombs, and whose only purpose is to emit a strong signal and to be an anchor. In these cases, the nodes will know the Euclidean distance to the anchors.

\section{Routing}
Our objective is to adapt geographic routing techniques that use Euclidean coordinates $(x,y)$ so that they work using the anchors coordinates.

\subsection{Greedy routing}
The greedy routing scheme consist in, given a current node $u$ and a destination $v$, choosing among the neighbors of $v$ the one which is the closest to the destination. If we use the greedy routing scheme to route a message using a distance function, the message will reach the destination given that there are no local minimum. This is easy to do with multidimensional coordinates.
Hence it is possible to directly use any geographic routing algorithm using greedy routing scheme.

\subsection{ROAM}
ROAM is a geographic routing algorithm based on the greedy routing scheme. It has been introduced in~\cite{HJM+09}. Briefly, it consists in marking nodes leading to local minima at which the greedy routing scheme would fail. Thus by using only unmarked nodes, it routes messages to the destination with the greedy routing scheme in a straightforward manner.

Hence, we can adapt ROAM so that it uses the anchors coordinates. We call the modified algorithm VROAM and it works as follows.

\begin{algorithm}
\caption{VROAM}

\begin{algorithmic}
\STATE Input : A node, a message and a destination.
\STATE Output : A node to which to forward the message.
\STATE The current node proceeds to a dead-end evaluation process;
\IF {The current node is not marked} 
\STATE it chooses the next node using a greedy routing scheme among its unmarked neighbors;
\ELSE 
\STATE it chooses the next node using an escape mode.
\ENDIF 
\end{algorithmic}
\end{algorithm}

\paragraph{dead-end evaluation process}
This process aims at marking nodes leading to a local minima. The current node checks if there are nodes closer to the destination than itself that are not already marked. If not, it marks itself and all its neighbors recursively execute a dead-end process. A node which is not marked as a dead-end during the process advertises itself as a possible exit to its neighbors. A node which is marked and does not know a possible exit and receives such a message stores the sender as a possible exit and recursively advertises itself as a possible exit.

\paragraph{escape mode}
In escape mode, a marked node forwards the message to a node which is a possible exit.

In Section~\ref{sec:simul}, we compare the efficiency of VROAM against ROAM.

\subsection{GRIC}
If it is possible to adapt algorithms based on the greedy routing scheme, we also want to adapt other algorithms. Since the nodes of the network are still in a surface, even with the new coordinate system, it is possible to use this surface and to define rotation. Using this, we can define a variant of GRIC using the anchors coordinates, we call it VGRIC.

\paragraph{How GRIC works}
For more details, refer to the original paper~\cite{PS07}. GRIC aims at sending a message towards an ideal direction $\did$ computed from the direction of the last hop  $\dlast$ and the direction of the destination $\ddir$. If the previous direction $\dlast$ was almost the direction of the destination $\ddir$, then $\did=\ddir$. Otherwise, $\did$ is computed by rotating $\dlast$, consistently with the previous hops (to ensure that while going around an obstacle we do not change the way we go around it).

\paragraph{VGRIC}
This protocol is an adaptation of GRIC so that it uses the anchors coordinates. We use an oriented plane tangent to the surface on which we make the rotation. In VGRIC, this plane is defined by two vectors \north and \east.

\begin{algorithm}
\caption{VGRIC}

\begin{algorithmic}
  \STATE Input : A node and the destination of its message $v_{dest}$, \dlast, \peast.
\STATE Output : A node to which to forward the message, \east.
\STATE Computes \north and \east
\STATE Computes \did
\STATE Choose the best neighbor
\end{algorithmic}
\end{algorithm}

\paragraph{Computation of the vectors}
We set $\vGamma=\{\frac{\overrightarrow{uv}}{d(u,v)}:v \in \Gamma(u)\}$  the set of normalized vectors defined using the current node and its neighbors.

We set $\ddir=\overrightarrow{uv_{dest}}$, and we choose among \vGamma the vector $\overrightarrow{i}$ minimising the scalar product (taking the absolute value) with \ddir. We then set $\north=\ddir.\dlast*\dlast+\ddir.\overrightarrow{i}*\overrightarrow{i}$ and $\east=\ddir.\overrightarrow{i}*\dlast-\ddir.\dlast*\overrightarrow{i}$

We then normalize both \north and \east and check if $\east.\peast \geq 0$. If not, we negate \east.

\paragraph{Computation of the ideal direction}
We use the previous direction $\dlast$. We set $x=\dlast.\east$ and $y=\dlast.\north$. If $x<0$, we negate it $x=-x$. We then make a rotation of the vector $x*\east + y*\north$ in the plane defined by $\north$ and $\east$ of a predefined angle $\alpha$ (we took $\alpha=0.3\pi$ in our experiments):
$\did=(\cos(\alpha) * y + \sin(\alpha) * x)*\north+(\cos(\alpha) * x - \sin(\alpha) * y)*\east$. If the rotation was to important, i.e. $\did.\east<0$ while $\did.\north>0$, then we set $\did=\north$.

\paragraph{Best neighbor} It is the neighbor used to define $\overrightarrow{i} \in \vGamma$ which maximizes $\overrightarrow{i}.\did$.

\section{Simulations}\label{sec:simul}

We experimented the proposed techniques for both ROAM and GRIC. The implementations were made using AlgoSensim~\cite{AlgoSensim}. We compare the results with the traditional ROAM and GRIC implementations.

\begin{figure}[tb]
 \begin{center}
  \includegraphics[width=5cm]{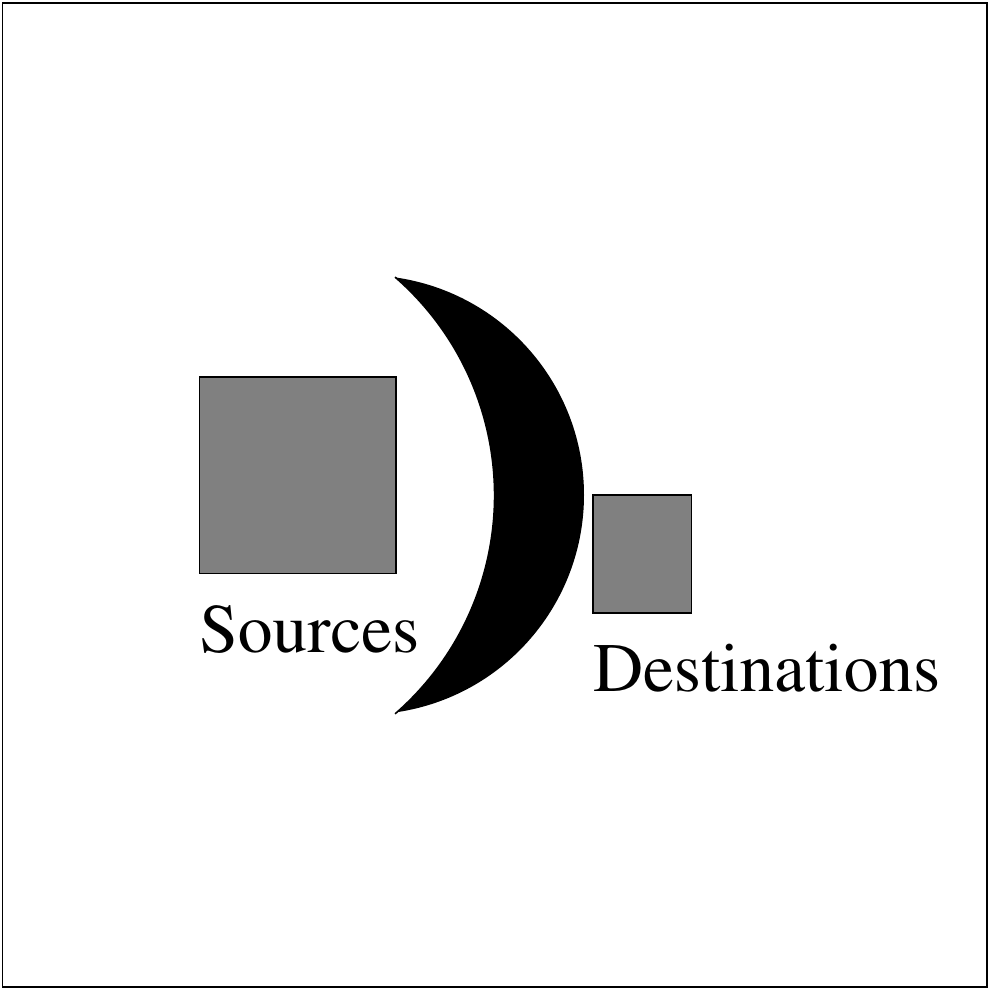}
  \caption{Simulation settings.}
  \label{fig:settings}
 \end{center}
\end{figure}

\paragraph{Settings}
We tested the techniques with six anchors, positioned at random. The distance to the anchors is either the exact distance (VROAM and VGRIC) or the distance in terms of hop (VROAMhop and VGRIChop). The scenario used is the following: a $50 * 50$ area is covered by 2000 to 8000 nodes which have communication radius 2. An obstacle with crescent shape is positioned in between a zone from which are emitted the messages and a zone towards which the messages are aimed, as displayed on Figure~\ref{fig:settings}.

\paragraph{Results}

\begin{figure}[tb]
 \begin{center}\hfill
  \includegraphics[width=5cm]{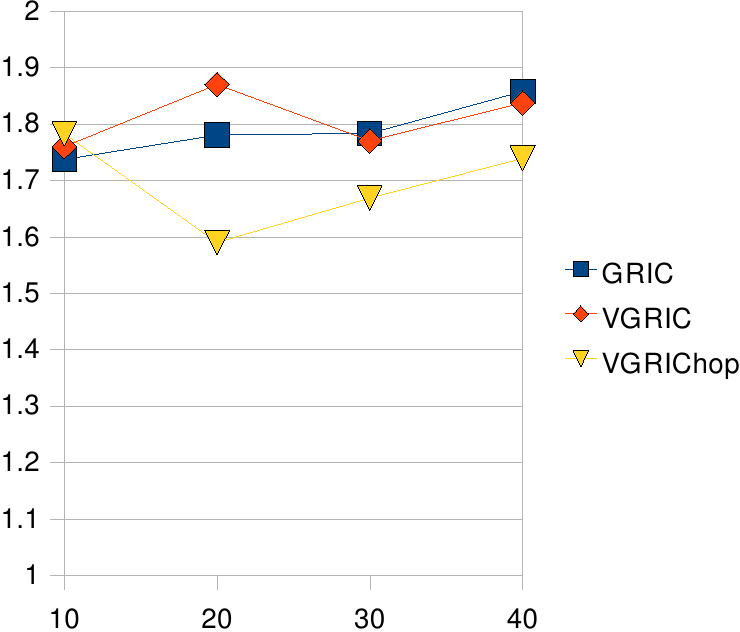}\hfill
  \includegraphics[width=5cm]{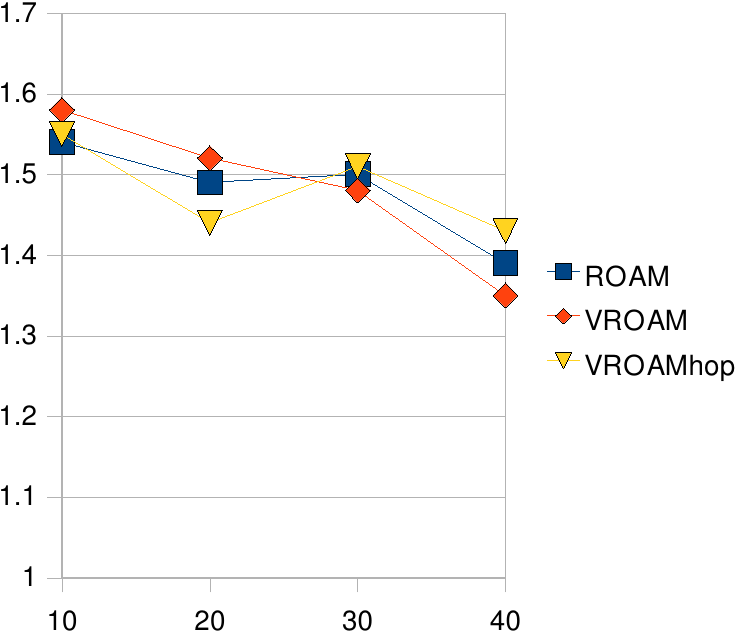}\hfill
    \caption{Simulation results: average stretch of paths function of average number of neighbors.}
  \label{fig:results}
 \end{center}
\end{figure}

Concerning ROAM, the experiments show that the use of raw coordinates to run routing algorithms give more than satisfactory results. Indeed the two modified algorithms perform as well as the original algorithm using real 2D coordinates. The path found by the three variants of ROAM have similar stretch. However, whereas ROAM and VROAM mark dead-end zones of similar surface areas, VROAMhop's dead-end zones tend to be smaller.

In \cite{RPSS03}, the authors argue that using the distance in terms of hops to the anchors often yields better results than using Euclidean distance. We share this opinion: indeed, the hop-count contains in itself information about the topology of the network, as it increases if there is an obstacle between a node and an anchor. This is supported by the experimental results concerning the size of dead-end zones. 

The fact that the three variants of ROAM compute paths of similar length does not contradict this. Indeed ROAM marks node leading to local minima and hence obtains information similar to the hop-count. This explains that all three variants perform as efficiently.

Concerning GRIC, with the Euclidean distance to the anchors, the routing paths are just slightly longer in average than with the normal coordinates, whereas with the hop distance, they are slightly shorter. Again, this is due to the fact that the hop distance contains more information than the Euclidean distance. Indeed, it is influenced by the obstacles and helps in avoiding them.

\section{Conclusion}
We have seen that routing directly using as coordinates the distance to anchors gives very good results and avoids heavy computations. The overhead is completely avoided in some scenarios, for instance when the distance to anchors is known through signal bombs. To further weaken localization assumptions, nodes may know their hop distance to only a subset of all anchors. This may correspond to the case when nodes are inside a building and hence cannot receive the signal from an anchor in a different room. In any case, it would imply to select wisely a subset of anchors, on which to route the message at each step.
Finally another planed development is to test this technique for routing algorithms with mobile nodes.

\nocite{*}
\bibliographystyle{alpha}
\bibliography{virtualRouting}
\label{sec:biblio}

\end{document}